\documentclass[ap,iop]{emulateapj}  

\usepackage{graphicx} 
\usepackage{apjfonts} 

\shorttitle{Pervasive linear polarization signals in the quiet Sun}
\shortauthors{Bellot Rubio \& Orozco Su\'arez}
\journalinfo{The Astrophysical Journal, 757:19 (7pp), 2012 September 20}
\submitted{Received 2012 April 13; accepted 2012 July 3; published 2012 August 31}

\begin{document}

\title{Pervasive linear polarization signals in the quiet Sun}
\author{L.~R.~Bellot Rubio$^1$}
\author{D.~Orozco Su\'arez$^2$}

\affil{$^1$ Instituto de Astrof\'{\i}sica de Andaluc\'{\i}a (CSIC),
  Apdo.\ 3004, 18080 Granada, Spain; lbellot@iaa.es}
\affil{$^2$ National Astronomical Observatory of Japan, 2-21-1 Osawa, 
Mitaka, Tokyo 181-8588, Japan}

\begin{abstract} 
This paper investigates the distribution of linear polarization
signals in the quiet-Sun internetwork using ultra-deep
spectropolarimetric data. We reduce the noise of the observations as
much as is feasible by adding single-slit measurements of the
Zeeman-sensitive \ion{Fe}{1} 630~nm lines taken by the {\em Hinode}
spectropolarimeter.  The integrated Stokes spectra are employed to
determine the fraction of the field of view covered by linear
polarization signals. We find that up to 69\% of the quiet solar
surface at disk center shows Stokes Q or U profiles with amplitudes
larger than 0.032\% (4.5 times the noise level of $7 \times 10^{-5}$
reached by the longer integrations).  The mere presence of linear
polarization in most of the quiet Sun implies that the weak
internetwork fields must be highly inclined, but we quantify this by
inverting those pixels with Stokes Q or U signals well above the
noise. This allows for a precise determination of the field
inclination, field strength, and field azimuth because the information
carried by all four Stokes spectra is used simultaneously. The
inversion is performed for 53\% of the observed field of view at a
noise level of $1.3 \times 10^{-4}
\, I_{\rm c}$. The derived magnetic distributions are thus
representative of more than half of the quiet-Sun internetwork.  Our
results confirm the conclusions drawn from previous analyses using
mainly Stokes I and V: internetwork fields are very inclined, but
except in azimuth they do not seem to be isotropically distributed.


\end{abstract}
\vspace*{1em}
\keywords{magnetic fields -- polarization -- Sun: photosphere --  Sun: surface magnetism}

\section{Introduction}

There is growing evidence that quiet-Sun regions outside of the
network---the so-called internetwork---are completely covered by
magnetic fields. These fields appear to be weak and highly inclined to
the vertical \citep[e.g.,][]{Orozco_apj2, Lites2,
2009A&A...495..607I,2009ApJ...701.1032A,2011A&A...527A..29B}, but
their characterization is challenging because they do not produce
linear polarization signals of sufficient amplitude as to be
measurable with the noise levels of current observations. Thus,
important parameters such as the field inclination have been inferred
rather indirectly, using Stokes I and V for the most part. While
Stokes I and V are sensitive to the field inclination
\citep{2010ApJ...711..312D} and additional constraints can be obtained
from the absence of Stokes Q and U, the fact is that without linear
polarization the determination of the field inclination is difficult
and subject to uncertainties (\citealt{2009ApJ...701.1032A};
cf.~\citealt{2012ApJ...751....2O}).

To improve the inferences one has to analyze both circular and linear
polarization signals at the same time. Using visible lines, however,
Stokes Q or U are detected only in a small fraction of the surface
area covered by the internetwork. For instance, the normal map and
deep-mode {\em Hinode} observations of
\cite{Lites2} show clear linear signals (4.5 times the noise level or
larger) in only 2\% and 27\% of the field of view, respectively
\citep[see][]{2012ApJ...751....2O}.  Even the more sensitive
ground-based \ion{Fe}{1} 630~nm measurements of
\citet{2008A&A...477..953M} do not go beyond 20\%. The rest of the
internetwork remains uncharted in linear polarization and may exhibit
different magnetic properties.

The lack of detection of linear signals is due to the weakness of the
fields. In the weak field regime, Stokes Q and U scale with the field
strength $B$ as $B^2$ and Stokes V as $B$, so weak fields generate
linear polarization much less efficiently than circular
polarization. Fortunately, Q and U are never zero except when the
field is vertical, which opens the door to their detection by
decreasing the noise of the measurements.  Since internetwork fields
appear to be highly inclined, linear polarization should exist almost
everywhere in the quiet Sun, albeit with low amplitudes.

Here, we push the capabilities of current instruments to a limit in
order to verify this conjecture. First, we combine the deep-mode
observations of \cite{Lites2} to improve their signal-to-noise ratio
by up to a factor of 11. Once the noise is reduced, the measurements
reveal clear linear polarization signals in most of the quiet-Sun
internetwork. This corroborates the results of earlier analyses and
puts them on a firmer basis. Second, we analyze the newly detected
signals to determine the properties of internetwork fields very
precisely. The derived properties apply to more than half of the
internetwork surface area.

\begin{figure*}[t]
\begin{center}
\resizebox{1\hsize}{!}{\includegraphics[]{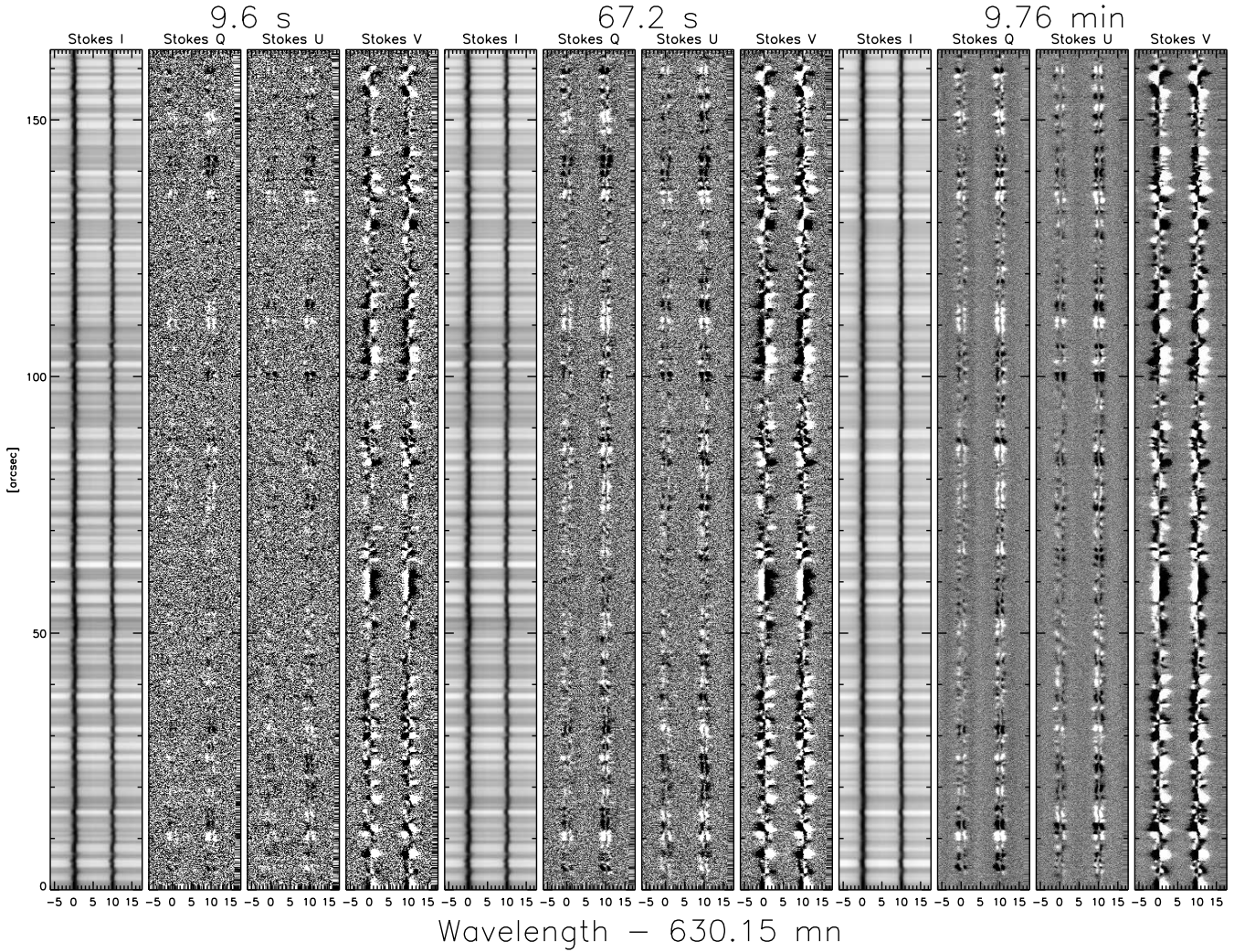}}
\end{center}
\vspace*{-1em}
\caption{Stokes I, Q, U and V spectra along the slit for integration 
times of 9.6~s (left), 67.2~s (middle), and 9.8~minutes (right). The
linear polarization signals (Stokes Q and U) stand out more
prominently all over the slit as the integration time increases, i.e.,
as the noise decreases. The spatial resolution is degraded only
slightly, as evidenced by the small changes in the intensity spectra.}
\vspace*{1em}
\label{fig1}
\end{figure*}

\section{Observations}

We use deep-mode observations taken by the spectropolarimeter
\citep[SP;][]{2001ASPC..236...33L} of the Solar Optical Telescope
\citep{2008SoPh..249..167T, 2008SoPh..249..197S, 2008SoPh..249..233I,
2008SoPh..249..221S} aboard {\em Hinode} \citep{2007SoPh..243....3K}.  On
2007 February 7, the {\em Hinode} SP measured the full Stokes vector of the
\ion{Fe}{1} 630~nm lines with an exposure time of 9.6~s and a pixel
size of 0\farcs16. The 160\arcsec\/ long slit of the SP was kept fixed
at disk center, sampling a very quiet region of the solar surface for
111 minutes.  These observations have a noise level of $8.0 \times
10^{-4}$ in units of the continuum intensity $I_{\rm c}$.

To reduce the noise down to $3.0 \times 10^{-4} \, I_{\rm c}$,
\citet{Lites2} added seven consecutive 9.6~s measurements, increasing 
the effective exposure time of the spectra to 67.2~s. After
integration, 27\% of the field of view showed Stokes Q or U signals
above 4.5 times the noise level, a threshold considered sufficient for
meaningful analysis of the Stokes vector through Milne--Eddington
inversions \citep{2012ApJ...751....2O}. This percentage is much larger
than the 2\% found in {\em Hinode}/SP normal maps recorded with exposure
times of 4.8~s and noise levels of $10^{-3} \, I_{\rm c}$
\citep{2012ApJ...751....2O}, but still insufficient to fully characterize 
the quiet-Sun internetwork.

\section{Data Analysis}

Here, we exploit the idea of \citet{Lites2} and continue the
integration of the signals to decrease the noise level even
further. By binning consecutive slits, we produce Stokes spectra with
effective exposure times ranging from 9.6~s (1~slit---the original
measurement) to 25.4~minutes (159 slits).  Figure~\ref{fig1} shows an
example of the signal-to-noise ratio gain in all the Stokes
parameters, for integrations of 9.6~s, 67.2~s, and 9.8~minutes. One
can clearly see that most of the polarization signals detected in the
less noisy spectra (those on the right) are also present in the other
panels, but virtually hidden by the larger noise. Thus, increasing the
exposure time brings to view very weak signals which could not be
observed before.

\begin{figure}[t]
\begin{center}
\resizebox{1\hsize}{!}{\includegraphics{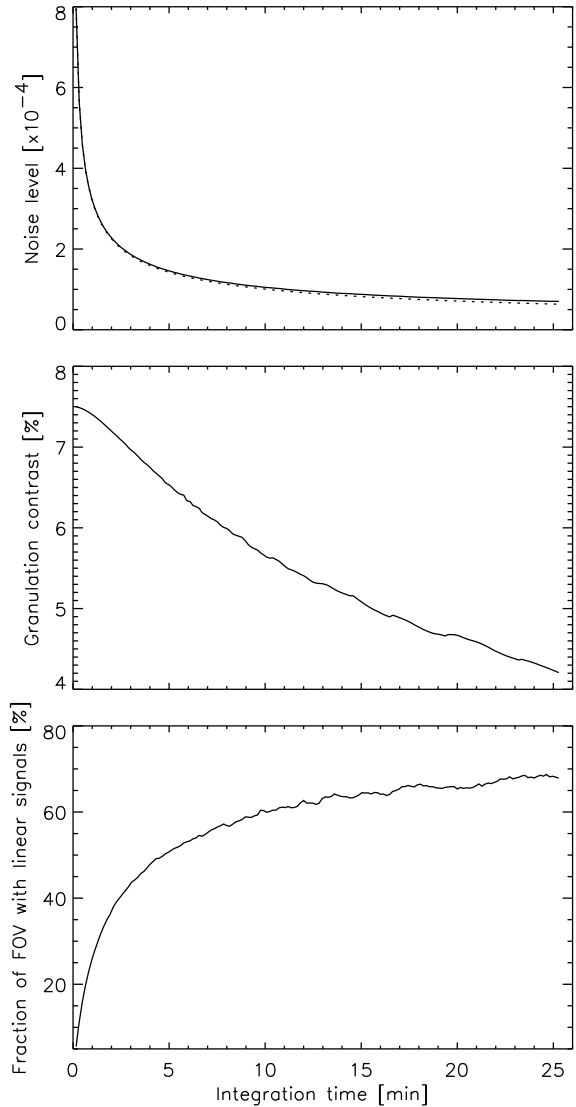}}
\end{center}
\caption{Top: noise level of the spectra vs.\ integration 
time (solid line). The noise is evaluated in the continuum of
Stokes Q and U. The dotted line shows the reduction expected for
photon noise, which is inversely proportional to the square root of
the integration time. Middle: rms continuum contrast vs.\ integration
time. Bottom: percentage of pixels with Stokes Q or U amplitudes
above 4.5 times the noise level vs.\ integration time. }
\label{fig2}
\end{figure}

The variation of the noise with time is shown in the top panel of
Figure~\ref{fig2} (solid line). After 25 minutes of integration, the
noise reaches an extremely low $7 \times 10^{-5} \, I_{\rm c}$ without
the help of any filtering. The curve follows very closely the behavior
expected for pure photon noise, indicating that other sources of
uncertainty (e.g., CCD readout noise) are less important. 

As the
effective exposure time is increased, however, the spatial resolution
of the spectra decreases due to changes in the solar scenery. The mean
lifetime of granular cells is of the order of 5--15 minutes
\citep{1987A&A...174..275A, 1989ApJ...336..475T, 1999ApJ...515..441H}
and magnetic flux concentrations, with similar lifetimes\footnote{ The
typical lifetime of the more horizontal internetwork fields is 1--10
minutes
\citep{2009A&A...495..607I,2010ApJ...723L.149D}. The more vertical 
internetwork fields live for 1.5--15 minutes, with a median value of
7.1 minutes \citep{milan}.}, are in constant motion buffeted by
convective flows. Owing to the dynamical evolution of the solar
surface, the superb spatial resolution of the {\em Hinode} SP is
progressively degraded. We quantify this degradation through the rms
contrast of the granulation in continuum intensity. The middle panel
of Figure~\ref{fig2} shows the variation of the contrast with
integration time. There is a steady decrease from 7.5\% for
integrations of 9.6~s to 4.2\% for integrations of 25~minutes, which
is significantly longer than the typical granular lifetime. Still,
these contrasts are higher than those obtained in ground-based
observations with shorter exposure times\footnote{For example, the
best granular contrast achieved with the Advanced Stokes Polarimeter
during moments of excellent seeing was 3.5\%
\citep{1996SoPh..163..223L}. With the help of adaptive optics, the 
Diffraction-Limited Spectropolarimeter reaches values of 5.2\% in
exposures of 4.3~s \citep{2004ApJ...613..600L}.  The 27~s long
observations of the Polarimetric Littrow Spectrograph analyzed by
\cite{2005A&A...436L..27K} had an rms contrast of 2.7\%.},
which testifies to the stability of the instrument and the advantages
of seeing-free observations from space.

\begin{figure*}[t]
\begin{center}
\resizebox{.7\hsize}{!}{\includegraphics{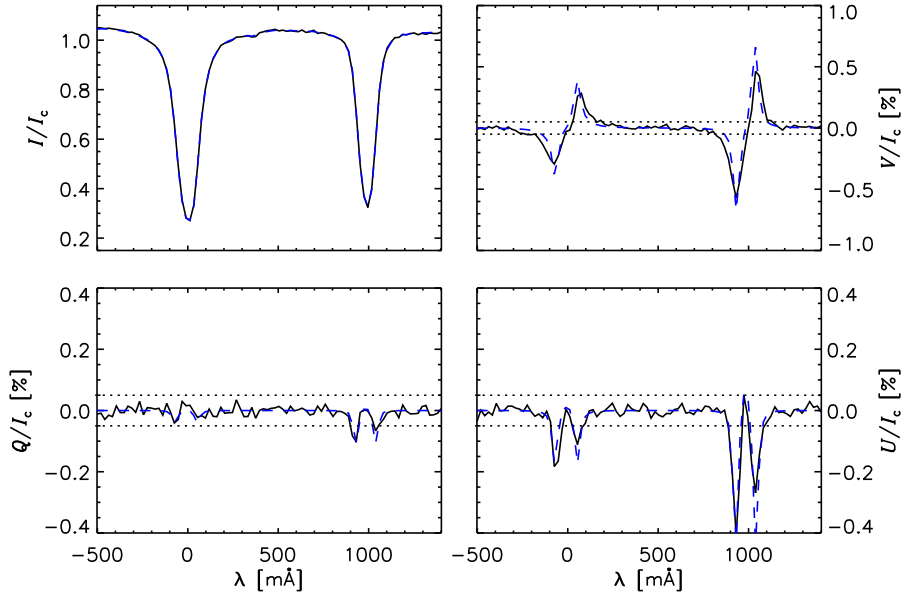}}
\end{center}
\caption{Example of observed Stokes profiles (solid) and best fit
resulting from a Milne--Eddington inversion (dashed). The observed
profiles correspond to an integration of 6.1 minutes and show Stokes Q
signals just above 4.5 times the noise level.}
\vspace*{.5em}
\label{fig3}
\end{figure*}

The benefits of reduced noise levels are immediately apparent from the
bottom panel of Figure~\ref{fig2}. Shown there is the percentage of
the field of view where we detect linear polarization (Stokes Q or U)
above 4.5 times the corresponding noise level. The fraction of surface
area covered by measurable linear polarization signals increases from
about 5\% to 69\% as the effective exposure time varies from 9.6~s to
25~minutes. However, it is not necessary to go that far to see linear
polarization in most of the internetwork: with integrations of
9.8~minutes, 60\% of the pixels---nearly two-thirds of the total---already
exhibit large Stokes Q and/or U amplitudes. The signals in the other
pixels do not exceed $4.5\sigma$, but most of them are real as
evidenced by their similar appearance in the two spectral lines and
the fact that they often show up in Stokes Q and U simultaneously.

From this we conclude that the solar internetwork is pervaded by 
linear polarization signals. 

\section{Magnetic Properties of the Internetwork}

\begin{figure*}[p]
\begin{center}
\resizebox{.95\hsize}{!}{\includegraphics[bb=5 140 465 510]{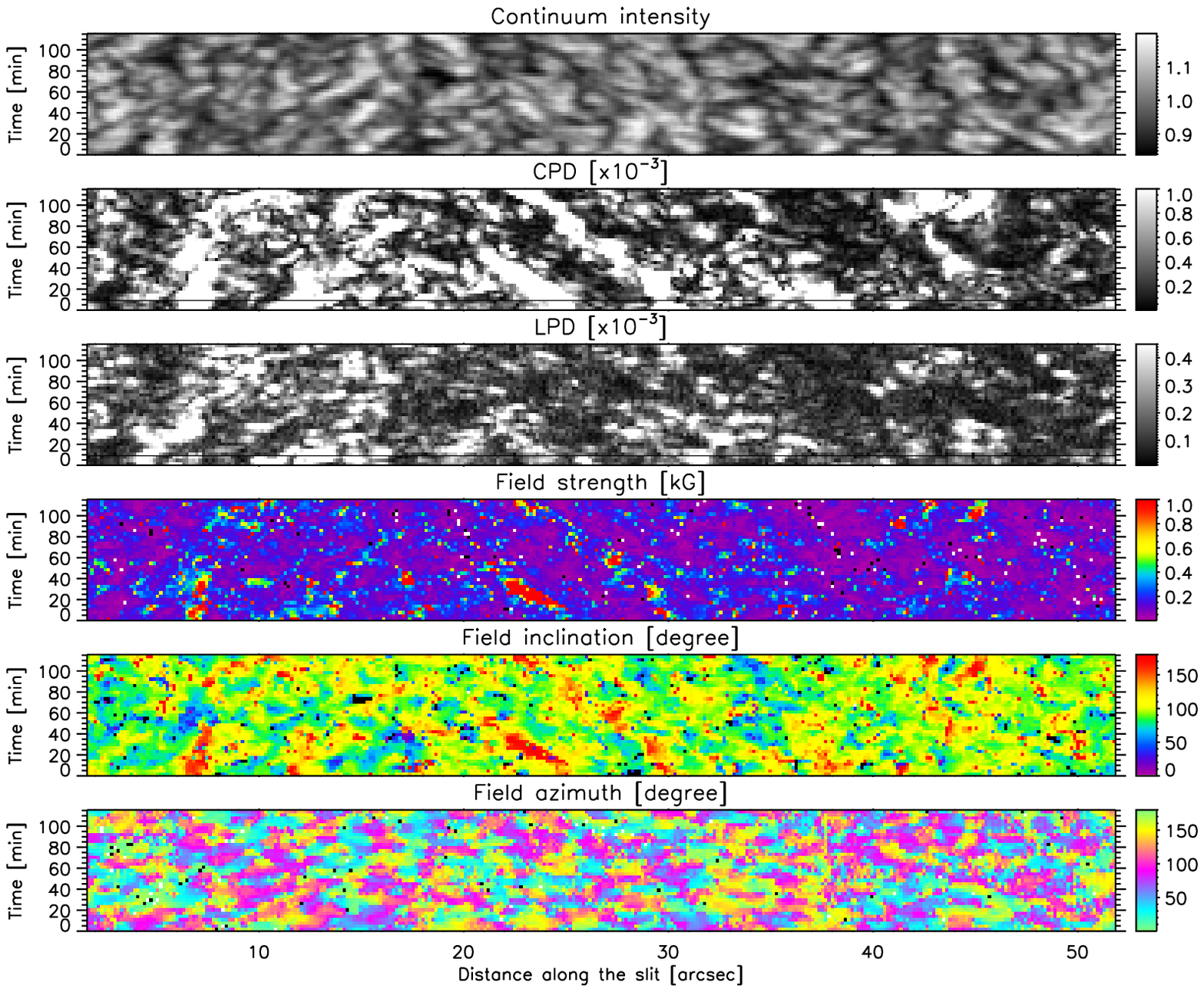}}
\end{center}
\caption{From top to bottom: temporal evolution of the observed 
continuum intensity, mean circular polarization degree, mean linear
polarization degree, and field strengths, inclinations, and azimuths
determined from the inversion.  The horizontal and vertical directions
represent distance along the slit and time (increasing from bottom to
top), respectively. Only a fraction of the total slit length is shown
to emphasize the spatial and temporal coherence of the signals. For
display purposes, the Stokes spectra used to construct this figure
were obtained through a running average of 6.1 minutes shifted by 3
minutes.
The mean circular and linear polarization degrees have been computed
using the two lines as $[\sum_{i=1}^N |V(\lambda_i)|/I(\lambda_i)]/N$
and $\sum_{i=1}^N \sqrt{Q(\lambda_i)^2+
U(\lambda_i)^2}/I(\lambda_i)]/N$, respectively.  $N$ is the number of
wavelength samples. The thin horizontal lines in the circular and
linear polarization maps mark the position of the cuts shown in
Figure~\ref{fig4b}. 
}
\label{fig4}
\end{figure*}

\begin{figure*}[p]
\begin{center}
\resizebox{.87\hsize}{!}{\includegraphics{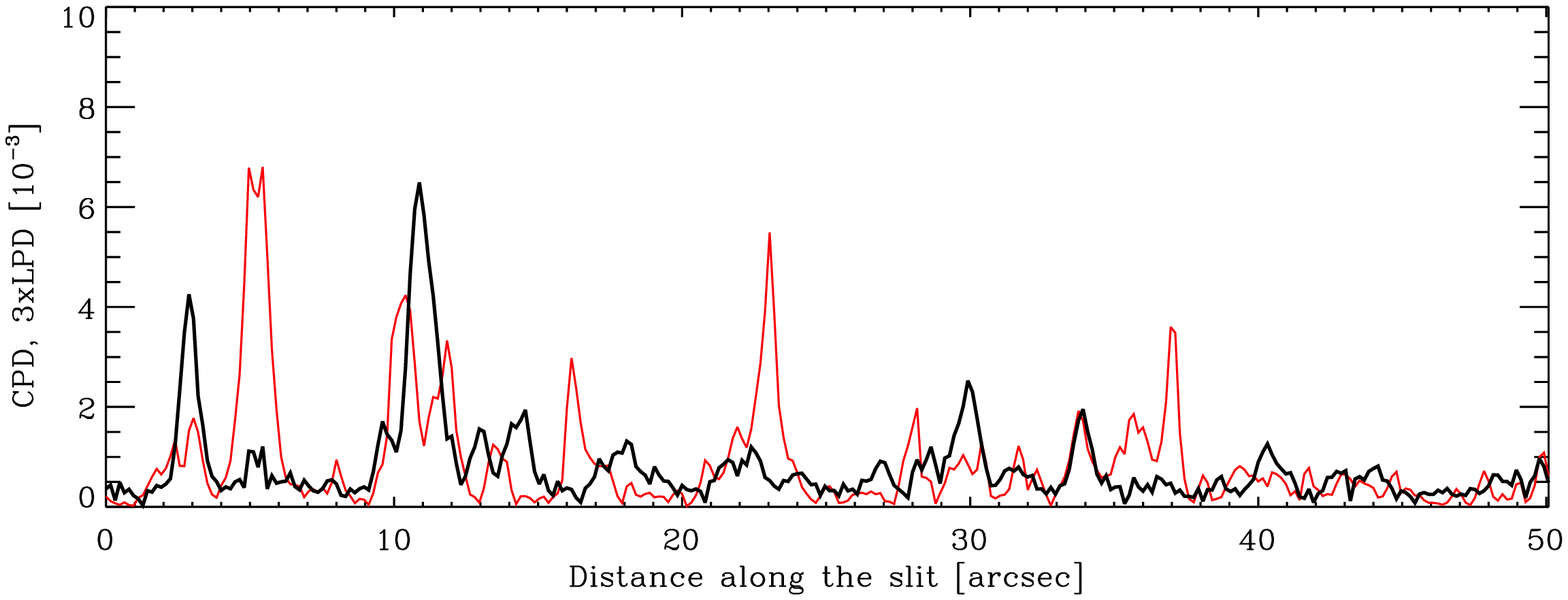}}
\end{center}
\vspace*{-2em}
\caption{Horizontal cuts of the circular and linear polarization 
maps shown in Figure~\ref{fig4} at $t=12$~minutes (thin and thick
lines, respectively). The exact position of the cuts is indicated by
horizontal black lines in the second and third panels of
Figure~\ref{fig4}. The linear polarization signal has been multiplied
by a factor of three for better visualization.  These cuts demonstrate
that the sizes of individual patches along the spatial direction are
comparable in linear and circular polarization.
}
\label{fig4b}
\end{figure*}

We take advantage of the large fraction of pixels showing clear Stokes
Q and U signals to derive the properties of internetwork fields with
high precision. In particular, we are interested in inferring their
strengths, inclinations, and azimuths. This is achieved by applying a
Milne--Eddington inversion to the observed spectra, as done by
\citet{Orozco_apj2} and others
\citep{Orozco_apj1,Orozco_pasj,2011A&A...527A..29B}. The inversion
determines nine free parameters characterizing the height-independent
properties of the magnetized atmosphere within the pixel, plus the
magnetic filling factor. The non-magnetic contribution in the pixel is
modeled using a local intensity profile \citep{Orozco_pasj}. According
to \citet{2012ApJ...751....2O}, this treatment of the non-magnetic 
component in single-slit observations may slightly overestimate the
field strength, but it does not affect the field inclination.

We choose to apply the inversion to the spectra integrated for
6.1~minutes. This is a compromise between high spatial resolution and
complete coverage of the solar internetwork. We could have inverted
the data with effective exposure times of 25~minutes, but at the cost
of reduced spatial resolution.  In order to maximize the chances that
the same pixel samples the same solar structure during the integration
time, we restrict ourselves to exposures of 6.1 minutes, which show
both high spatial resolution and large surface coverage: 53\% of the
pixels in the data set exhibit clear Stokes Q or U signals with
amplitudes larger than 4.5 times the noise level (of $1.3 \times
10^{-4} \, I_{\rm c}$), and 88\% have Stokes V signals above the same
threshold.  Thus, the analysis of these observations is representative
of more than half of the internetwork surface area, preserving most of
the spatial resolution offered by {\em Hinode}.

As an example, Figure~\ref{fig3} displays the inversion results for 
a pixel with Stokes Q signals just above the $4.5\sigma$ criterion. Note
the extremely low noise level and the quality of the fit, which
captures the essential properties of the observed spectra. In this
case, the inversion indicates a magnetic field strength of 200~G, a
field inclination of $103^\circ$, an azimuth of $130^\circ$, and a
magnetic filling factor of 0.2. The Milne--Eddington inversion cannot
reproduce the small asymmetries of the observed profiles because of
the assumption of height-independent atmospheric parameters.  As
pointed out by \citet{2011A&A...530A..14V}, most of the profiles
recorded by the {\em Hinode} SP are asymmetric to a larger or smaller
degree.  The asymmetries carry a great deal of information, whose
extraction requires more complex inversion schemes
\citep[e.g.,][]{2008ApJ...674..596S, 2011A&A...526A..60V, 
2012ApJ...748...38S}. For the exploratory purposes of this study,
however, a Milne--Eddington inversion is sufficient.

Figure~\ref{fig4} displays the time sequence of continuum intensity,
circular, and linear polarization degrees for the 6.1 minute
integration, as well as the field strengths, inclinations, and
azimuths resulting from the inversion. The horizontal axis represents
distance along the slit, while the vertical axis represents time (from
bottom to top).  Each row within the panels corresponds to an
integration of 6.1 minutes. Despite the long effective exposure time,
one can clearly distinguish small-scale flux concentrations which
evolve in a coherent and smooth manner.  For example, the proper
motions of the structures produce inclined streaks in the panels.
Interestingly, the circular and linear polarization maps show patches
of comparable size in the horizontal direction (the spatial
direction), all over the slit. Figure 5 illustrates this result. A
different behavior was seen in the normal maps of the {\em Hinode} SP
and the filtergrams acquired by IMaX
\citep{2011SoPh..268...57M}, the vector magnetograph of SUNRISE
\citep{2010ApJ...723L.127S, 2011SoPh..268....1B}.  Those instruments
suggested a quiet-Sun internetwork consisting of relatively large
circular polarization patches covering most of the surface and
smaller, transient linear polarization patches that appear and
disappear continually
\citep{Lites2,2010ApJ...723L.149D}. The long integrations displayed in
Figure~\ref{fig4} uncover much weaker signals, with the result that
the size of the linear polarization patches increases (probably
because their outer parts start to be detectable now).

Most of the structures seen in circular and linear polarization are
recognizable in the field strength, inclination, and azimuth maps of
Figure~\ref{fig4}. Only the largest circular signals happen to be
associated with fields stronger than 500~G, which are also the most
vertical ones. The linear signals come from weak and very inclined
fields. It is interesting to note that the azimuths show a smooth
pattern in both spatial and temporal directions. The existence of
structure in the azimuth map demonstrates the very precise
determination of the three magnetic field components allowed by the
small noise of the linear polarization profiles. 

\begin{figure}[t]
\begin{center}
\resizebox{.94\hsize}{!}{\includegraphics{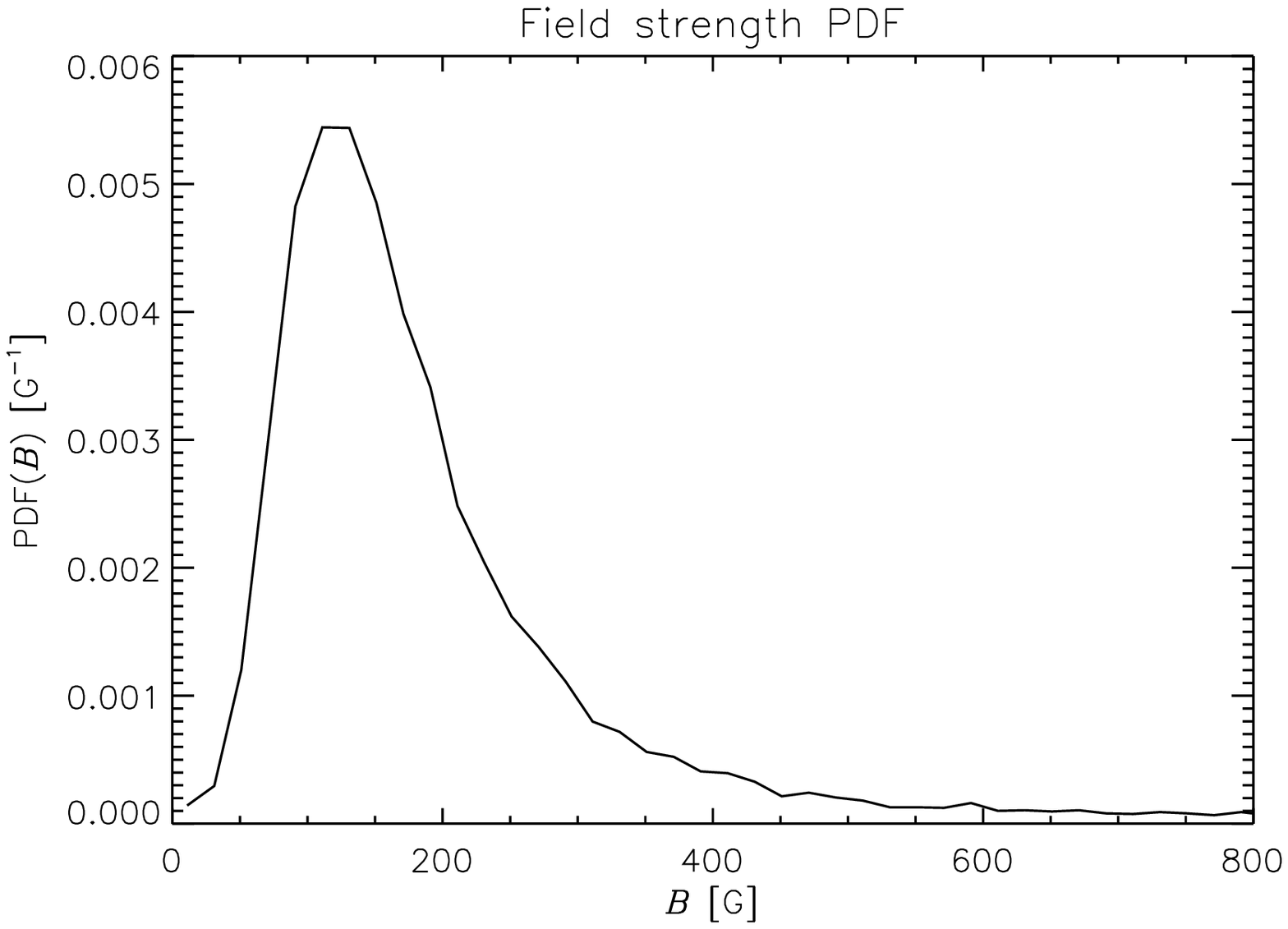}}
\resizebox{.94\hsize}{!}{\includegraphics{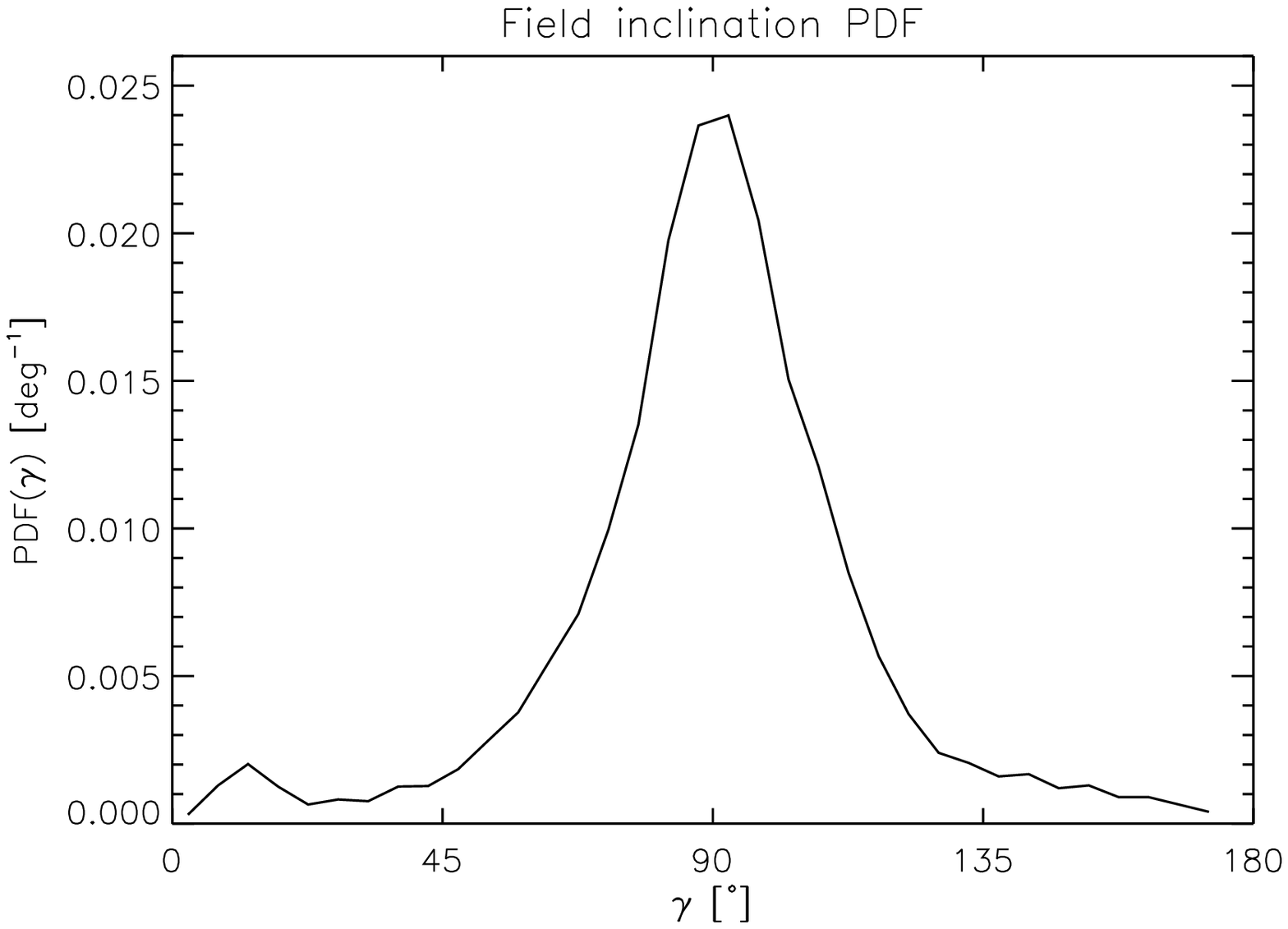}}
\resizebox{.94\hsize}{!}{\includegraphics{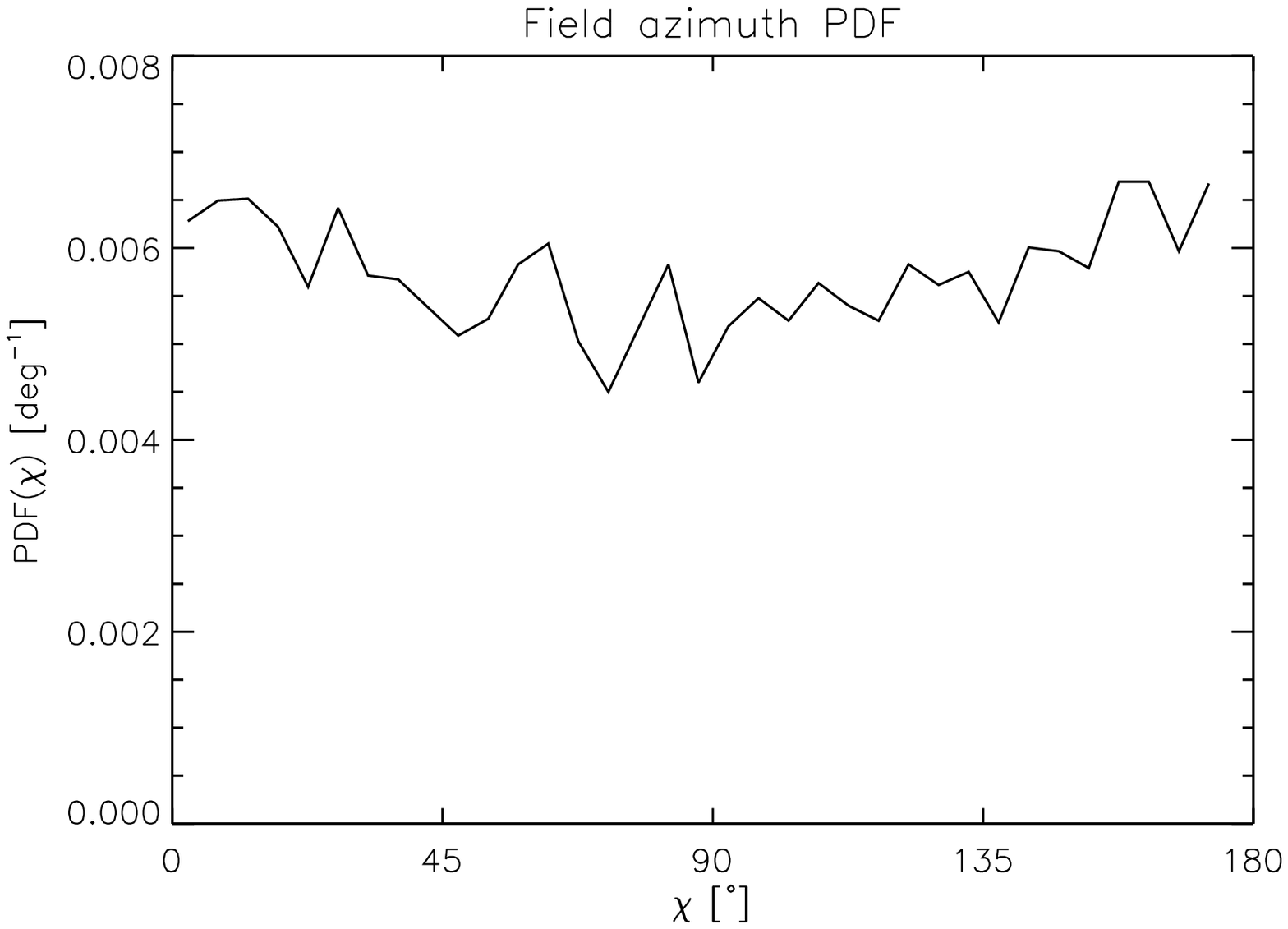}}
\end{center}
\vspace*{-1em}
\caption{Distributions of magnetic field strengths, inclinations, 
and azimuths resulting from the inversion of pixels with linear
polarization (Stokes Q or U) signals larger than 0.059\% (4.5 times
the noise level) in the ultra-deep SP maps averaged over
6.1~minutes. These PDFs account for 53\% of the observed field of
view.}
\vspace*{1em}
\label{fig5}
\end{figure}

To obtain the distribution of field strengths, inclinations, and
azimuths we only consider the 10,419 pixels having Stokes Q or U
amplitudes larger than 4.5 times the noise level\footnote{By this
choice we ensure that the maximum possible information is used to
infer the vector magnetic field. Nearly 100\% of the 10,419 pixels
with significant linear polarization signals also have Stokes V
signals above the $4.5\sigma$ threshold, meaning that four non-zero
Stokes parameters are contributing to the determination of the vector
magnetic field. The converse is not true: 39\% of the pixels showing
Stokes V amplitudes larger than $4.5\sigma$ do not have linear
polarization signals above $4.5\sigma$. Since the inversion of these
pixels might not be as accurate as that of the other pixels, we
refrain from including them in the analysis. We note, however, that
the results derived from all the pixels with Stokes Q, U {\em or} V
signals above $4.5\sigma$ are fully compatible with the ones presented
in this paper, in particular the field distributions of
Figure~\ref{fig5}: the inclination distribution still shows a
pronounced---albeit slightly smaller---peak at 90$^{\circ}$, and the
field strength distribution is shifted to weaker fields by about
20~G. Those results account for 92.2\% of the area covered by the
internetwork.}. They  account for 53\% of the quiet-Sun
internetwork. The noise of $1.3
\times 10^{-4} \, I_{\rm c}$ reached by the 6.1 minute integration is
equivalent to a polarimetric sensitivity of 0.2~G in longitudinal
field and 15~G in transverse field\footnote{These values have been
obtained as the strength of the vertical (horizontal) fields that
produce circular (linear) polarization signals of $1\sigma$ with the
mean thermodynamic parameters of the quiet Sun.}.

Figure~\ref{fig5} displays the resulting probability density functions
(PDFs). As can be seen in the top panel, the field strength
distribution peaks at 110~G, with a fast decrease toward stronger
fields. No hump is detected at kG fields. The distribution also
decreases rapidly toward 0~G. This appears to be a robust feature of
the inversion, not caused by the noise of the profiles or the
selection criteria\footnote{The $4.5\sigma$ polarization threshold
adopted for Stokes Q and U corresponds to a transverse field of 32~G,
which is very far from the maximum of the distribution at 110~G. For
this reason, we consider it unlikely that the peak is an artifact of
the selection criteria. }.

The central panel of Figure~\ref{fig5} reveals an inclination
distribution dominated by horizontal fields (inclinations are measured
from the local vertical). The PDF has a sharp maximum at $90^\circ$,
with tails toward more vertical fields decreasing very quickly.  Thus,
in addition to purely horizontal fields, one also finds inclined
fields in the solar internetwork. Consistent with earlier claims
(e.g.,
\citealt{2012ApJ...751....2O}; Borrero \& Kobel, in preparation), the
distribution of inclinations deviates significantly from a cosine
function, and therefore does not appear to be isotropic.

Finally, the PDF of field azimuth shows a random distribution of
orientations in the solar surface. 

\section{Discussion and Conclusions}
By pushing the integration time to a limit, the spectropolarimetric
measurements taken by the {\em Hinode} SP reveal a quiet-Sun
internetwork full of linear polarization signals. At least 69\% of the
area occupied by the internetwork shows unambiguous Stokes Q or U
signals in the \ion{Fe}{1} 630~nm lines when the noise is decreased
down to $7\times10^{-5} \, I_{\rm c}$. This, however, is a difficult
endeavor, requiring integrations of about 25 minutes which degrade the
spatial resolution of the observations. With integrations of 10
minutes (comparable to the granular lifetime), the degradation is not
so severe and 60\% of the internetwork still shows measurable linear
polarization. Similar coverage levels have been reported earlier from
near-infrared \ion{Fe}{1} 1565~nm measurements, but at lower spatial
resolution
\citep{2008A&A...477..953M,2009A&A...502..969B}.

The origin of these signals is unclear as yet. They are pervasive, so
the mechanism producing them must operate all over the solar surface.
The observations are compatible with the existence of an ``ocean'' of
weak and highly inclined fields which are present everywhere and can
only be detected when the noise is reduced below a certain level---a
true photospheric magnetic carpet. At the same time, it is possible
that part of the observed Stokes Q and U signals come from transient
{\em Horizontal Internetwork Fields} (HIFs;
\citealt{1996ApJ...460.1019L}, \citealt{2009A&A...495..607I},
\citealt{2010ApJ...723L.149D}) and small-scale magnetic loops that
emerge in the solar photosphere and progress into higher layers or
disappear there \citep{2007ApJ...666L.137C, 2009ApJ...700.1391M,
2010A&A...511A..14G, 2010ApJ...713.1310I, 2010ApJ...714L..94M}. In
fact, short-lived, localized HIFs emerging at high rates can be
expected to completely fill the slit with signals for long enough
integrations (a similar conclusion was reached by
\citealt{2010ApJ...718L.171I} from the analysis of deep vector
magnetograms taken with the {\em Hinode} Narrowband Filter Imager).  The
only way to distinguish between scenarios is to investigate the time
evolution of the fields at high cadence and very low noise
levels. This task should be performed by the new generation of large
($>$ 1.5m) telescopes, including NST, GREGOR, and, on the longer term,
also ATST, EST, and Solar-C.

We have used the deep {\em Hinode} integrations to derive the properties of
internetwork fields through a Milne--Eddington inversion with constant
atmospheric parameters and a local non-magnetic component.  To avoid
excessive degradation of the spatial resolution and large variations
of the solar conditions,  we use Stokes spectra integrated for 6.1
minutes, comparable to the mean granular lifetime.  Only pixels
showing Stokes Q or U signals above 4.5 times the noise level ($1.3
\times 10^{-4} \, I_{\rm c}$) were considered for analysis. They
represent 53\% of the solar internetwork. These pixels also have
significant circular polarization signals, so the magnetic field
parameters derived from the inversion are very accurate because all
the available information is used at the same time
\citep{2009ApJ...701.1032A,2011A&A...527A..29B}. Our threshold 
of 4.5 times the noise level is rather strict, but designed to pick
only real signals. The probability that a random variable following a
normal distribution of zero mean and standard deviation $\sigma$
exceeds $4.5\sigma$ is $7 \times 10^{-6}$. Thus, the probability that
a pixel without any linear polarization is selected erroneously would
be $7 \times 10^{-6} \times 90 \times 2 = 0.0013$, since 90
wavelengths are used to determine the maximum of the polarization
profile and there are 2 of them (Stokes Q and U). With such a low
probability, one can be completely sure that the inverted profiles are
real and not artifacts of the noise.

The magnetic field distributions inferred from the inversion confirm
that internetwork fields are weak and very inclined. The peaks of field
strength and inclination occur at 110~G and $90^\circ$, respectively,
as had been estimated from noisier observations in the past
\citep{Orozco_apj2, 2012ApJ...751....2O}. The field inclination
distribution is not isotropic and does not consist only of purely
horizontal fields, since there exist a non-negligible amount of pixels
with inclinations between $45^\circ$ and $90^\circ$ (or between
$90^\circ$ and $135^\circ$).  Finally, the field azimuth appears to be
randomly oriented, as required to explain the absence of Stokes U
signals in Hanle-effect measurements of spectral lines near the limb
\citep[e.g.,][]{1982SoPh...80..209S}.

Because of their very similar properties (weak strengths near or below
the Hanle saturation limit, large inclinations, random orientation of
azimuths, and almost complete coverage of the solar surface), it is
tempting to speculate that the magnetic fields revealed by the long
integrations of {\em Hinode} are those sampled by Hanle-depolarization
measurements. In the future, progress will come from the observation
of the temporal evolution of internetwork fields, including their
emergence and disappearance in the solar atmosphere. With current
instruments, such observations are not feasible due to insufficient
photon collecting power. This limitation highlights the need for
larger telescopes, both on the ground and in space.

\acknowledgments

This work has been funded by the Spanish MICINN through projects
AYA2011-29833-C06-04 and PCI2006-A7-0624, and by Junta de
Andaluc\'{\i}a through project P07-TEP-2687, including a percentage
from European FEDER funds. D.O.S.  thanks the Japan Society for the
Promotion of Science (JSPS) for financial support through its
postdoctoral fellowship program for foreign researchers. {\em Hinode} is a
Japanese mission developed and launched by ISAS/JAXA, with NAOJ as
domestic partner and NASA and STFC (UK) as international partners. It
is operated by these agencies in co-operation with ESA and NSC
(Norway). The use of NASA's Astrophysical Data System is gratefully
acknowledged.


\begin{thebibliography}{}
\bibitem[Alissandrakis et 
al.(1987)]{1987A&A...174..275A} Alissandrakis, C.~E., Dialetis, D., \& Tsiropoula, G.\ 1987, \aap, 174, 275 

\bibitem[Asensio Ramos(2009)]{2009ApJ...701.1032A} Asensio Ramos, A.\ 2009, 
\apj, 701, 1032 

\bibitem[Barthol et al.(2011)]{2011SoPh..268....1B} Barthol, P., Gandorfer, 
A., Solanki, S.~K., et al.\ 2011, \solphys, 268, 1 


\bibitem[Beck 
\& Rezaei(2009)]{2009A&A...502..969B} Beck, C., \& Rezaei, R.\ 2009, \aap, 502, 969 

\bibitem[Borrero 
\& Kobel(2011)]{2011A&A...527A..29B} Borrero, J.~M., \& Kobel, P.\ 2011, \aap, 527, A29

\bibitem[Centeno et al.(2007)]{2007ApJ...666L.137C} Centeno, R., 
Socas-Navarro, H., Lites, B., et al.\ 2007, \apjl, 666, L137 

\bibitem[G{\"o}m{\"o}ry et 
al.(2010)]{2010A&A...511A..14G} G{\"o}m{\"o}ry, P., Beck, C., Balthasar, H., et al.\ 2010, \aap, 511, A14 


\bibitem[Danilovic et al.(2010)]{2010ApJ...723L.149D} Danilovic, S., Beeck, 
B., Pietarila, A., et al.\ 2010, \apjl, 723, L149 


\bibitem[del Toro Iniesta et al.(2010)]{2010ApJ...711..312D} del Toro 
Iniesta, J.~C., Orozco Su{\'a}rez, D., \& Bellot Rubio, L.~R.\ 2010, \apj, 711, 312 

\bibitem[Go{\v{s}}i{\'c}(2012)]{milan} Go\v{s}i\'c, M. 2012, Master 
Thesis, University of Granada (Spain)

\bibitem[Hirzberger et al.(1999)]{1999ApJ...515..441H} Hirzberger, J., 
Bonet, J.~A., V{\'a}zquez, M., \& Hanslmeier, A.\ 1999, \apj, 515, 441

\bibitem[Ichimoto et al.(2008)]{2008SoPh..249..233I} Ichimoto, K., Lites, 
B., Elmore, D., et al.\ 2008, \solphys, 249, 233 

\bibitem[Ishikawa 
\& Tsuneta(2009)]{2009A&A...495..607I} Ishikawa, R., \& Tsuneta, S.\ 2009, \aap, 495, 607 


\bibitem[Ishikawa 
\& Tsuneta(2010)]{2010ApJ...718L.171I} Ishikawa, R., \& Tsuneta, S.\ 2010, \apjl, 718, L171 


\bibitem[Ishikawa et al.(2010)]{2010ApJ...713.1310I} Ishikawa, R., Tsuneta, 
S., \& Jur{\v c}{\'a}k, J.\ 2010, \apj, 713, 1310 



\bibitem[Khomenko et 
al.(2005)]{2005A&A...436L..27K} Khomenko, E.~V., Mart{\'{\i}}nez Gonz{\'a}lez, M.~J., Collados, M., et al.\ 2005, \aap, 436, L27 



\bibitem[Kosugi et al.(2007)]{2007SoPh..243....3K} Kosugi, T., Matsuzaki, K., Sakao, T., et al.\ 2007, \solphys, 243, 3 


\bibitem[Lites(1996)]{1996SoPh..163..223L} Lites, B.~W.\ 1996, \solphys, 
163, 223 

\bibitem[Lites 
\& Socas-Navarro(2004)]{2004ApJ...613..600L} Lites, B.~W., \& Socas-Navarro, H.\ 2004, \apj, 613, 600

\bibitem[Lites et al.(2001)]{2001ASPC..236...33L} Lites, B.~W., Elmore, D.~F., \& Streander, K.~V.\ 2001, in ASP Conf. Proc., Vol. 236, Advanced Solar Polarimetry - Theory, Observation, and Instrumentation, ed.\ M. Sigwarth (San Francisco, CA: ASP), 33

\bibitem[Lites et al.(2008)]{Lites2} Lites, B.~W., Kubo, M., Socas-Navarro, H., et al.\ 2008, \apj, 672, 1237

\bibitem[Lites et al.(1996)]{1996ApJ...460.1019L} Lites, B.~W., Leka, 
K.~D., Skumanich, A., Martinez Pillet, V., 
\& Shimizu, T.\ 1996, \apj, 460, 1019 

\bibitem[Mart{\'{\i}}nez Gonz{\'a}lez et 
al.(2008)]{2008A&A...477..953M} Mart{\'{\i}}nez Gonz{\'a}lez, M.~J., 
Collados, M., Ruiz Cobo, B., \& Beck, C.\ 2008, \aap, 477, 953 


\bibitem[Mart{\'{\i}}nez Gonz{\'a}lez 
\& Bellot Rubio(2009)]{2009ApJ...700.1391M} Mart{\'{\i}}nez Gonz{\'a}lez, M.~J., \& Bellot Rubio, L.~R.\ 2009, \apj, 700, 1391

\bibitem[Mart{\'{\i}}nez Gonz{\'a}lez et al.(2010)]{2010ApJ...714L..94M} 
Mart{\'{\i}}nez Gonz{\'a}lez, M.~J., Manso Sainz, R., Asensio Ramos, A., 
\& Bellot Rubio, L.~R.\ 2010, \apjl, 714, L94 


\bibitem[Mart{\'{\i}}nez Pillet et al.(2011)]{2011SoPh..268...57M} 
Mart{\'{\i}}nez Pillet, V., et al.\ 2011, \solphys, 268, 57 

\bibitem[Orozco Su{\'a}rez et al.(2007a)]{Orozco_apj1} Orozco 
Su{\'a}rez, D., Bellot Rubio, L.~R., 
\& del Toro Iniesta, J.~C.\ 2007a, \apjl, 662, L31 

\bibitem[Orozco Su{\'a}rez et al.(2007b)]{Orozco_apj2} Orozco Su{\'a}rez, D., Bellot Rubio, L.~R., \& del Toro Iniesta, J.~C.\ et al.\ 2007b, \apjl, 670, L61 

\bibitem[Orozco Su{\'a}rez et al.(2007c)]{Orozco_pasj} Orozco Su{\'a}rez, D., Bellot Rubio, L.~R., \& del Toro Iniesta, J.~C.\ et al.\ 2007c, \pasj, 59, 837 


\bibitem[Orozco Su{\'a}rez 
\& Bellot Rubio(2012)]{2012ApJ...751....2O} Orozco Su{\'a}rez, D., \& Bellot Rubio, L.~R.\ 2012, \apj, 751, 2 


\bibitem[Sainz Dalda et al.(2012)]{2012ApJ...748...38S} Sainz Dalda, A., 
Mart{\'{\i}}nez-Sykora, J., Bellot Rubio, L., 
\& Title, A.\ 2012, \apj, 748, 38

\bibitem[Shimizu et al.(2008)]{2008SoPh..249..221S} Shimizu, T., Nagata, 
S., Tsuneta, S., et al.\ 2008, \solphys, 249, 221


\bibitem[Socas-Navarro et al.(2008)]{2008ApJ...674..596S} Socas-Navarro, 
H., Borrero, J.~M., Asensio Ramos, A., et al.\ 2008, \apj, 674, 596 

\bibitem[Solanki et al.(2010)]{2010ApJ...723L.127S} Solanki, S.~K., et al.\ 
2010, \apjl, 723, L127 


\bibitem[Stenflo(1982)]{1982SoPh...80..209S} Stenflo, J.~O.\ 1982, 
\solphys, 80, 209 

\bibitem[Suematsu et al.(2008)]{2008SoPh..249..197S} Suematsu, Y., Tsuneta, 
S., Ichimoto, K., et al.\ 2008, \solphys, 249, 197 


\bibitem[Title et al.(1989)]{1989ApJ...336..475T} Title, A.~M., Tarbell, 
T.~D., Topka, K.~P., et al.\ 1989, \apj, 336, 475

\bibitem[Tsuneta et al.(2008)]{2008SoPh..249..167T} Tsuneta, S., Ichimoto, K., Katsukawa, Y., et al.\ 2008, \solphys, 249, 167 

\bibitem[Viticchi{\'e} et 
al.(2011)]{2011A&A...526A..60V} Viticchi{\'e}, B., S{\'a}nchez Almeida, J., Del Moro, D., \& Berrilli, F.\ 2011, \aap, 526, A60 


\bibitem[Viticchi{\'e} 
\& S{\'a}nchez Almeida(2011)]{2011A&A...530A..14V} Viticchi{\'e}, B., \& S{\'a}nchez Almeida, J.\ 2011, \aap, 530, A14 



\end{thebibliography}
\end{document}